\documentclass[useAMS,usenatbib,usegraphicx]{mn2e}

\title[$z=7.3$ LAEs behind Lensing Clusters]{A Search for $z=7.3$ Ly$\alpha$ Emitters behind Gravitationally Lensing Clusters}
\author[Kazuaki Ota, Johan Richard, Masanori Iye, Takatoshi Shibuya, Eiichi Egami and Nobunari Kashikawa]{Kazuaki Ota$^{1}$\thanks{E-mail:
otakz@kusastro.kyoto-u.ac.jp}, Johan Richard$^{2}$, Masanori Iye$^{3,4,5}$, Takatoshi Shibuya$^{5}$,
\newauthor 
Eiichi Egami$^{6}$ and Nobunari Kashikawa$^{3,5}$\\
$^{1}$Department of Astronomy, Kyoto University, Kitashirakawa-Oiwake-cho, Sakyo-ku, Kyoto 606-8502, Japan\\
$^{2}$CRAL, Observatoire de Lyon, Universit$\acute{e}$ Lyon 1, 9 Avenue Ch. Andr$\acute{e}$, 69561 Saint Genis Laval Cedex, France\\
$^{3}$National Astronomical Observatory of Japan, 2-21-1 Osawa, Mitaka, Tokyo, 181-8588, Japan\\
$^{4}$Department of Astronomy, Graduate School of Science, University of Tokyo, 7-3-1 Hongo, Bunkyo-ku, Tokyo 113-0033, Japan\\
$^{5}$The Graduate University for Advanced Studies, 2-21-1 Osawa, Mitaka, Tokyo, 181-8588, Japan\\
$^{6}$Steward Observatory, University of Arizona, 933 N. Cherry Avenue, Tucson, AZ 85721, USA}

\begin{document}

\date{Accepted 2011 July XX. Received 2011 July XX; in original form 2011 July XX}

\pagerange{\pageref{firstpage}--\pageref{lastpage}} \pubyear{2002}

\maketitle

\label{firstpage}

\begin{abstract}
We searched for $z=7.3$ Ly$\alpha$ emitters (LAEs) behind two gravitationally lensing clusters, Abell 2390 and CL 0024, using the Subaru Telescope Suprime-Cam and a narrowband filter NB1006 ($\lambda_c \sim 1005$ nm, FWHM $\sim21$ nm). The combination of the fully depleted CCDs of the Suprime-Cam, sensitive to $z\sim 7$ Ly$\alpha$ emission at $\sim 1$ $\mu$m, and the magnification by the lensing clusters can be potentially a powerful tool to detect faint distant LAEs. Using the NB1006 and deep optical to mid-infrared images of the clusters taken with the {\it Hubble} and {\it Spitzer Space Telescopes}, we investigated if there exist objects consistent with the color of $z=7.3$ LAEs behind the clusters. We could not detect any LAEs to the unlensed Ly$\alpha$ line flux limit of $F_{{\rm Ly}\alpha} \simeq 6.9 \times 10^{-18}$ erg s$^{-1}$ cm$^{-2}$. Using several $z=7$ Ly$\alpha$ luminosity functions (LFs) from the literatures, we estimated and compared the expected detection numbers of $z\sim7$ LAEs in lensing and blank field surveys in the case of using an 8m class ground based telescope. Given the steep bright-end slope of the LFs, when the detector field-of view (FOV) is comparable to the angular extent of a massive lensing cluster, imaging cluster(s) is more efficient in detecting $z\sim7$ LAEs than imaging a blank field.  However, the gain is expected to be modest, a factor of two at most and likely much less depending on the adopted LFs. The main advantage of lensing-cluster survey, therefore, remains to be the gain in depth and not necessarily in detection efficiency. For much larger detectors, the lensing effect becomes negligible and the efficiency of LAE detection is proportional to the instrumental FOV. We also investigated the NB1006 images of the three $z\sim7$ $z$-dropout galaxy candidates previously detected in Abell 2390 and found that none of them are detected in the NB1006. Two of them are consistent with the predictions from the previous studies that they would be at lower redshifts. The other one has a photometric redshift of $z\simeq 7.3$, and if we assume that it is at $z=7.3$, the unlensed Ly$\alpha$ line flux would be very faint: $F_{{\rm Ly}\alpha} < 4.4 \times 10^{-18}$ erg s$^{-1}$ cm$^{-2}$ ($1\sigma$ upper limit) or equivalent width of $W_{{\rm Ly}\alpha}^{\rm rest} < 26$\AA. Its Ly$\alpha$ emission might be attenuated by neutral hydrogen, as recent studies show that the fraction of Lyman break galaxies displaying strong Ly$\alpha$ emission is lower at $z\sim7$ than at $z\la 6$.
\end{abstract}


\section{Introduction}
Detecting and characterizing distant galaxies such as Lyman break galaxies (LBGs) and Ly$\alpha$ emitters (LAEs) at high redshift have been a major method to probe galaxy evolution and reionization in the early universe. In addition to the fact that $z \la 6$ LBGs are now routinely observed, the number of $z \ga 7$ LBG candidates has dramatically increased recently owning to the great sensitivity and wide area view of the {\it Hubble Space Telescope} ({\it HST}) WFC3 as well as new generations of wide near-infrared cameras on ground based facilities, and their physical and statistical properties have been intensively investigated \citep[e.g.,][]{Ouchi09,Bouwens10,Bunker10,Castellano10,Hickey10,Oesch10,Wilkins10,Grazian11,McLure11,Yan11}. On the other hand, the number of $z \ga 7$ LAE samples has also increased \citep{Ota08,Ota10,Hibon10,Hibon11,Hibon12,Shibuya11,Tilvi10,Krug12}, and some of LAEs even include spectroscopically identified objects from $z \ga 7$ LBG samples \citep{Fontana10,Lehnert10,Ono12,Pentericci11,Schenker12,Vanzella11}. LAEs at $z \ga 6$ have been used as a probe of reionization, and the significant variation in their Ly$\alpha$ luminosity function (LF) has been observed from $z\simeq6$ out to $z \simeq 8$, which might indicate either the evolution of LAEs, the possible suppression of Ly$\alpha$ emission by neutral hydrogen gas, or the combination of both effects \citep[e.g.,][]{Shimasaku06,Kashikawa06,Ota08,Ouchi08,Hibon10,Hu10,Nakamura11,Ota10,Ouchi10,Tilvi10,Clement11,Hibon11,Hibon12,Kashikawa11,Krug12,Shibuya11}. However, these results still remain tentative for $z > 7$ LAEs, because the number of candidates detected and studied to date is small. Thus, observing more and more $z>7$ LAEs and increasing the sample size are essential to drawing the statistically more robust conclusions about $z>7$ Ly$\alpha$ LF as well as comparing with the $z>7$ LBG studies.

LAEs at $z>7$ could be faint and difficult to detect due to the possible suppression of Ly$\alpha$ emission by neutral hydrogen, if the universe is not fully reionized. Using the magnification by gravitationally lensing clusters is one of the effective ways to overcome this problem, as many authors have already demonstrated its great potential for detecting faint high redshift galaxies \citep[e.g.,][]{Bouwens09,Bradley08,Franx97,Hu02,Kneib04,Richard08,Santos04,Stark07,Willis08}. The surface area and the flux of an image behind a lensing cluster in the source plane are both magnified in the image plane at the time of observation. This increases the chance to detect intrinsically very faint objects, if located behind the positions of large magnification in the lensing cluster. However, at the same time, the larger the magnification factor is, the smaller the intrinsic (or "unlensed") surface area is. This decreases the chance to detect faint rare objects. In fact, in the case of lensing surveys, the unlensed sky area probed tends to be small, such as a few to a few tens arcmin$^2$ or even smaller, depending on how many clusters are observed. Usually, the efficiency of galaxy survey is dependent on the balance between its area and depth as well as the slope of galaxy LF. It is suggested that if the slope of the LF, $-d( \log \phi ) / d (\log L)$, is steeper than $1$, the gain in depth makes up for the loss in area in the overall number of lensed galaxies to be detected \citep{Broadhurst95,Maizy10}. \citet{Bouwens09} actually assessed and confirmed this effect by conducting a search for $z \sim 7$ $z$-dropout galaxies behind 11 massive lensing clusters and finding 1--4 candidates. In the case of $z\ga7$ LAEs, the bright end slope of LF tends to be $\ga1.5$. For example, \citet{Kashikawa11} obtained $\log \phi^* = -3.28$ and $\log L^* = 42.76$ with fixed $\alpha = -1.5$ for the Ly$\alpha$ LF based on $\sim 50$ $z=6.6$ LAEs by fitting the \citet{Schechter76} LF. This best-fit LF has $-d( \log \phi ) / d (\log L) \ga 1.56$ at $L \ga L^*$. Similarly, the best-fit Schechter LFs for 265 and $\sim30$ $z=6.6$ LAEs from \citet{Ouchi10} and \citet{Hu10} have the slopes of $-d( \log \phi ) / d (\log L) \ga 1.55$ and $\ga 1.55$ at $L \ga L^*$, respectively. Moreover, \citet{Hibon10} obtained $\log \phi^* = -4.4$ to $-3.0$ and $\log L^* = 42.5$--$43.3$ with $\alpha = -1.5$ for the Ly$\alpha$ LF based on their $z=7.7$ LAE candidates, and the slope is $-d( \log \phi ) / d (\log L) \ga 1.55$ at $L \ga L^*$. These LF slopes indicate that even shallower imaging of lensing clusters could be potentially useful for detecting $z \sim 7$--8 $L \ga L^*$ LAEs.

\begin{figure}
\includegraphics[width=62mm,angle=-90]{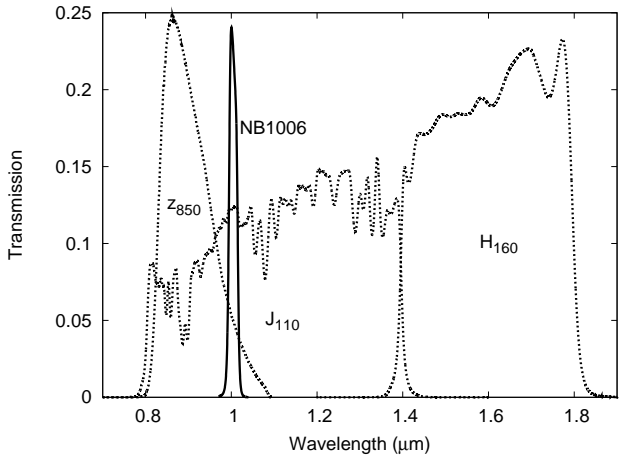}
\caption{Filter transmission of the Subaru Suprime-Cam narrowband NB1006 (solid curve) as well as the {\it HST} ACS and NICMOS broadbands $z_{850}$, $J_{110}$ and $H_{160}$ (dashed curves) used for our photometry.}
\label{FilterTransmission}
\end{figure}

Meanwhile, improving detector sensitivity is another complementary way to overcome the difficulty in detecting $z>7$ LAEs. In 2008, the CCDs of the Subaru Telescope Suprime-Cam were replaced with the new fully depleted CCDs to make it more sensitive to the red wavelength at $\ga 0.9$ $\mu$m and more advantageous for detecting $z \ga 7$ galaxies. This had pushed the maximum redshift of galaxies that the Suprime-Cam can probe from $z=7$ to $z=7.3$, and thus we made a new narrowband filter NB1006 (FWHM $\sim21$ nm centered at $\sim 1005$ nm) designed to target $z\sim 7.18$--7.35 Ly$\alpha$ emission. The transmission curve of the NB1006 is shown in the Figure \ref{FilterTransmission}. Motivated by the potential usefulness of the lensing magnification for $z>7$ LAE surveys, we conducted a search for $z=7.3$ LAEs behind two lensing clusters Abell 2390 and CL 0024 using the NB1006 combined with the fully depleted CCDs on the Suprime-Cam. In this paper, we present the results of this search. The paper is organized as follows. In Section 2, we describe the imaging observation of the lensing clusters and data reduction. Then, the photometry and $z=7.3$ LAE candidate selection are conducted in Section 3. In Section 4, we present the results and discuss their implications. We conclude and summarize our results in Section 5. Throughout, we adapt a concordance cosmology with $(\Omega_m, \Omega_{\Lambda}, h) = (0.3, 0.7, 0.7)$, and AB magnitudes, unless otherwise specified.

\section{Observation and Data}
\subsection{Optical to Mid-infrared Broadband Data}
The target clusters for our $z=7.3$ LAE lensing search are Abell 2390 ($z=0.228$) and CL 0024 ($z=0.390$). The deep optical, near-infrared and mid-infrared broadband images taken with the {\it HST} and the {\it Spitzer} as well as the Subaru Telescope are available for Abell 2390 and CL 0024. These data were obtained from observations and archives and reduced by \citet{Richard08}. The optical images of Abell 2390 were taken with the {\it HST} WFPC2 F555W, F814W and ACS F850LP filters (hereafter denoted as $V_{555}$, $I_{814}$ and $z_{850}$, respectively), while CL 0024 with the {\it HST} ACS F475W, F625W, F775W and F850LP filters (hereafter $g_{475}$, $r_{625}$, $i_{775}$ and $z_{850}$, respectively). Also, the near and mid-infrared images of the both clusters were taken with the {\it HST} NICMOS F110W, F160W, the Subaru MOIRCS $K$-band and the {\it Spitzer} IRAC 3.6 $\mu$m and 4.5 $\mu$m filters (hereafter $J_{110}$, $H_{160}$, $K$, [3.6] and [4.5], respectively). The depth of these optical and infrared images, taken from \citet{Richard08}, are listed in Table 1.

\subsection{Narrowband Observations and Data Reduction\label{NBreduction}}
The narrowband NB1006 imaging observations of the lensing clusters with the Subaru Suprime-Cam were carried out at the dark clear nights on 2009 October 15 and 16. At each night, they were observed by making use of the interval of two hours between the time right after the evening twilight and the time right before the start of our another observation, the NB1006 field search for $z=7.3$ LAEs. The result of the $z=7.3$ LAE field search is presented in \citet{Shibuya11} paper. We have chosen Abell 2390 and CL 0024 as targets because (1) they were observable at that time, (2) the deep optical to mid-infrared images taken with the {\it HST} and the {\it Spitzer} are available for them, and (3) \citet{Richard08} had detected a $z$-dropout, A2390-z2, whose photometric redshift has the probability peak at $z\sim 7.3$, suggesting that it may have $z\sim 7.3$ Ly$\alpha$ emission. The sky conditions during observations were photometric with a seeing of $\sim 0\farcs5$--$0\farcs7$. By taking the 20 minutes dithered exposure frames, we obtained a total of 2.0 and 1.3 hours of imaging data for Abell 2390 and CL 0024, respectively. 

\begin{figure}
\includegraphics[width=85mm,angle=0]{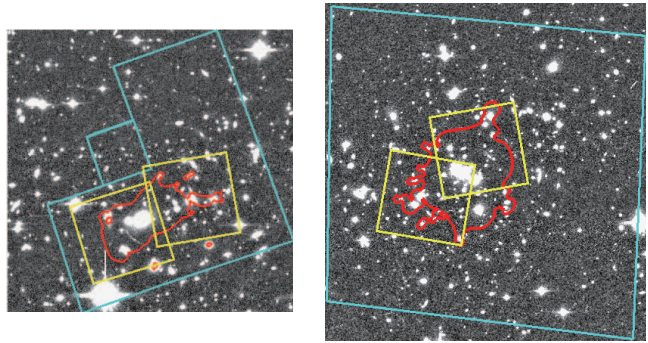}
\caption{({\it Left}) The NB1006 image of Abell 2390 cut into the $3\farcm2 \times 3\farcm2$ size (the NB1006-A2390 image in Section \ref{NBreduction}). ({\it Right}) The NB1006 image of CL 0024 cut into the $3\farcm6 \times 3\farcm8$ size (the NB1006-CL0024 image in Section \ref{NBreduction}). The north is up and the east to the left. The critical line for a $z=7.3$ source in each lensing cluster is shown by the red curve. The areas covered by the {\it HST} WFPC2 (the cyan line in the NB1006-A2390 image), ACS (the cyan line in the NB1006-CL0024 image) and NICMOS (the yellow line) are superposed on the images. As described in Sections \ref{SelectionCriteria} and \ref{CandidateSelection}, we search for $z=7.3$ LAEs in the region displayed with the cyan line by using the LAE selection criterion (1)--(3), and also apply the supplementary LAE selection criterion (4) with the NICMOS $J_{110}$ band to the objects only in the region displayed with the yellow line.}
\label{NB1006images}
\end{figure}

We have reduced the NB1006 image frames using the software SDFRED \citep{Yagi02,Ouchi04} in the same standard manner as in \citet{Kashikawa04} and \citet{Ota08}. The dithered NB1006 exposure frames were combined. The seeing sizes of the final combined NB1006 images of Abell 2390 and CL 0024 are $0\farcs73$ and $0\farcs83$, respectively. The spectrophotometric standard star GD71 \citep{Oke90} was imaged during the observations to calibrate the photometric zeropoints of these stacked images, which were turned out to be NB1006 $=$ 31.41 and 31.43 mag ADU$^{-1}$ for Abell 2390 and CL 0024, respectively. Abell 2390 and CL 0024 are located in the center of the large areas of NB1006 images taken with the Suprime-Cam (field-of-view is $34'\times27'$). Hence, we cut the NB1006 images into the smaller sizes around Abell 2390 and CL 0024, the similar sizes to the effective areas of the WFPC2 and ACS images of Abell 2390 and CL 0024, respectively. Hereafter, we call them NB1006-A2390 and NB1006-CL0024 images and use them for the subsequent photometry and analyses. These images are shown in Figure \ref{NB1006images}. The limiting magnitudes measured with a $2\farcs0$ diameter aperture in the NB1006-A2390 and NB1006-CL0024 images reached NB1006 $=23.78$ and 23.70 mag at $3 \sigma$ , respectively. The depth of these images are also shown in Table 1.

\begin{table*}
\centering
\begin{minipage}{178mm}
\begin{center}
\caption{Depth of the imaging data}
\begin{tabular}{@{}lcccccccccc@{}}
\hline 
Abel 2390 &           & $V_{555}$ & $I_{814}$ & $z_{850}$ & NB1006 & $J_{110}$ & $H_{160}$ & $K$   & [3.6] & [4.5]\\
\hline 
          &           & 26.6      & 26.2      & 26.82     & 23.78  & 26.27     & 26.54     & 25.6  & 23.9  & 23.9 \\    
\hline
CL 0024   & $g_{475}$ & $r_{625}$ & $i_{775}$ & $z_{850}$ & NB1006 & $J_{110}$ & $H_{160}$ & $K$   & [3.6] & [4.5]\\
\hline
          & 27.81     & 27.75     & 27.67     & 27.28     & 23.70  & 26.20     & 26.60     & 25.7  & 23.9  & 24.1 \\    
\hline
\end{tabular}\\
\end{center}
NOTES: The depth of NB1006 is $3\sigma$ while that of all the other bands are $5\sigma$ taken from \citet{Richard08}. The following diameter apertures were used for the photometry: $0\farcs3$ for the {\it HST} WFPC2 and ACS optical bands, $2\farcs0$ for the Subaru Suprime-Cam NB1006, $0\farcs6$ for the {\it HST} NICMOS $J_{110}$ and $H_{160}$ bands, $0\farcs5$ for the Subaru MOIRCS $K$-band and $3\farcs0$ for the {\it Spitzer} IRAC 3.6 and 4.5 $\mu$m bands.
\end{minipage}
\end{table*}

\section{Photometry and Candidate Selection}
\subsection{Photometry}
We conducted photometry to make the NB1006-detected object catalogs. First of all, the astrometry and the pixel scales of all the optical to mid-infrared broadband images were matched to those of the NB1006-A2390 and NB1006-CL0024 images. Also, the point spread functions (PSFs) of the $z_{850}$ images were convolved to match those of the NB1006-A2390, NB1006-CL0024 and $J_{110}$-band images, because we have to calculate the $z_{850} - {\rm NB1006}$ and $z_{850} - J_{110}$ colors by measuring the $z_{850}$, NB1006 and $J_{110}$ magnitudes using the same diameter aperture in order to select $z=7.3$ LAE candidates (See Section \ref{SelectionCriteria}). 

Then, source detection and photometry were carried out with the SExtractor software \citep{BA96}. The pixel size of the NB1006 images taken with the Suprime-Cam CCDs is $0\farcs202$ pixel$^{-1}$. We considered an area larger than five contiguous pixels with a flux greater than $2\sigma$ (i.e., $2 \times$ background rms) to be an object. Object detection was first made in the NB1006-A2390 and NB1006-CL0024 images, and then photometry was done in the images of other wavebands using the double-imaging mode. The aperture magnitudes of detected objects were measured with the {\tt MAG\_APER} parameter, and aperture corrections were applied to obtain the total magnitudes. As adopted by \citet{Richard08}, we also used the $0\farcs3$ and $0\farcs6$ diameter apertures and the corrections of 0.3 and 0.6 mag for the WFPC2/ACS and NICMOS images, respectively, for our photometry. These aperture corrections were estimated using bright isolated unsaturated stars and assuming a point source \citep{Richard08}. For the measurements of the $z_{850} - {\rm NB1006}$ and $z_{850} - J_{110}$ colors, we used $2\farcs0$ and $0\farcs6$ diameter apertures. We did not apply any aperture correction for NB1006 because we did not use total NB1006 magnitude but used only the $2''$ diameter aperture magnitude for measuring the color of $z_{850}-{\rm NB1006}$ and limiting magnitudes of the NB1006-A2390 and NB1006-CL0024 images. Finally, by combining the photometry in all the wavebands, we constructed the object catalogs for Abell 2390 and CL 0024.

\subsection{Detection Completeness}
\begin{figure}
\includegraphics[width=85mm]{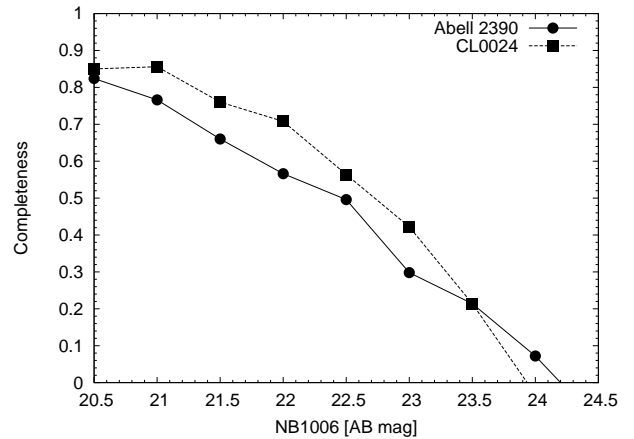}
\caption{Detection completeness of Abell 2390 and CL 0024 NB1006 images, calculated for every 0.5 mag bin. The completeness do not reach 1.0 even for the objects with bright magnitudes. This is because the blended or overlapped objects tend to be counted as one object by the SExtractor.}
\label{Completeness}
\end{figure}

What fraction of objects in an image we can reliably detect by photometry depends on the magnitudes and overlapping of objects. The fraction usually decreases as the magnitudes become fainter due to their faintness. Also, the detectability of target objects is affected by their overlapping with foreground objects. Especially, this effect could be large in the case of lensing clusters where many galaxies are clustered in a space. To examine what fraction of objects in the NB1006-A2390 and NB1006-CL0024 images SExtractor can detect or fails to detect to fainter magnitude, we measured the detection completeness of our photometry for those images. Using the IRAF task {\tt starlist}, we first created 5,000 artificial objects with random but uniform spatial and magnitude distributions, ranging from 20 to 27 mag. Then, we spread them over the NB1006-A2390 and NB1006-CL0024 images, by using the IRAF task {\tt mkobject}. SExtractor was run for source detection in exactly the same way as our actual photometry. Finally, we calculated the ratio of the number of detected artificial objects to that of created ones to obtain the detection completeness. We repeated this procedure five times and averaged the obtained completeness. The result is shown in Figure \ref{Completeness}. The 50\% completeness for NB1006-A2390 and NB1006-CL0024 images correspond to ${\rm NB1006}\sim 22.5$ and $\sim 22.75$, respectively. 

\begin{figure*}
\centering
\begin{minipage}{175mm}
\includegraphics[width=62mm,angle=-90]{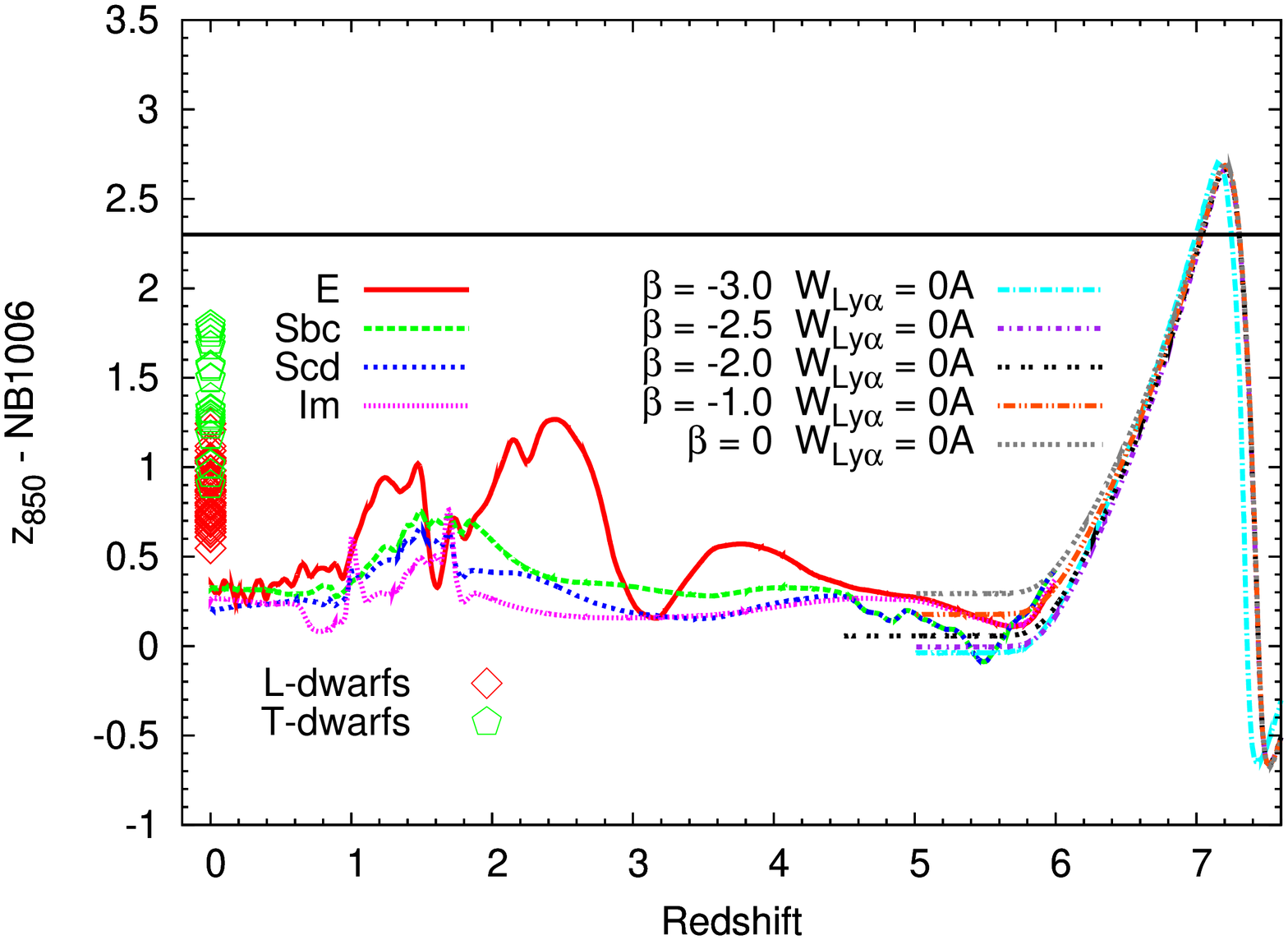}
\includegraphics[width=62mm,angle=-90]{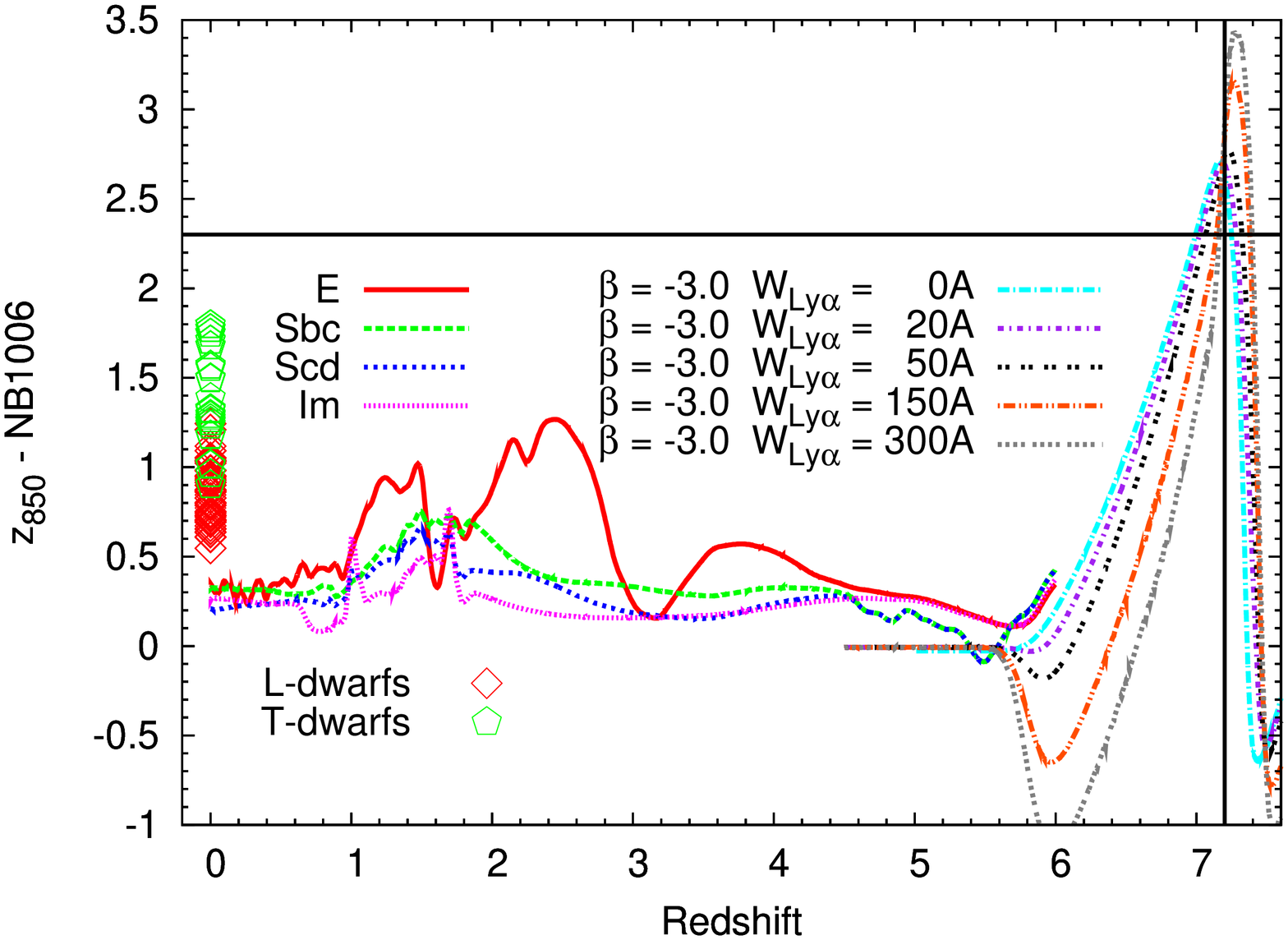}
\caption{$z_{850} - {\rm NB1006}$ color as a function of redshift for various types of galaxies. The colors of E (elliptical), Sbc, Scd and Im (irregular) galaxies are calculated by using \citet{CWW80} template spectra. The color of an LAE is calculated by using the spectrum $f_{\lambda} \propto \lambda^{\beta}$ for a variety of UV continuum slopes $\beta$ and Ly$\alpha$ equivalent widths $W_{{\rm Ly}\alpha}^{\rm rest}$. Also, colors of L1--L9.5 and T0.5--T8 type dwarf stars are calculated by using their spectra and plotted (see the text in Section \ref{SelectionCriteria} for the details of dwarf data). ({\it Left}) Colors of LBGs (spectra with $W_{{\rm Ly}\alpha}^{\rm rest}=0$). Though they do not have Ly$\alpha$ emission, LBGs show the narrowband excess of $z_{850} - {\rm NB1006} \sim 2.3$--2.7 at $z\sim7$--7.3, but amount of excess does not change much with $\beta$. At these redshifts, the UV continuum is detected only at the narrow wavelength range corresponding to the NB1006 passband and the red edge and low sensitivity portion of $z_{850}$-band. Hence, a difference in $\beta$ makes only a negligible change in the NB1006 and $z_{850}$ fluxes and thus amount of excess. ({\it Right}) Colors of an LBG and LAEs with $\beta=-3$ are plotted as a representative, as recent study of \citet{Ono10} suggests that the $z\sim7$ LBGs ($z_{850}$-dropouts) and $z=5.7$ and $z=6.6$ LAEs have the continuum slope of $\beta\simeq-3$ (See Figure \ref{CandidateColor3} for the case of $\beta=-2$). The LAE color clearly shows the narrowband excess at $7.2 \la z \la 7.4$. The horizontal and vertical lines indicate the NB1006-excess LAE selection criterion we adopt (See Section \ref{SelectionCriteria} for details).}
\label{CandidateColor1}
\end{minipage}
\end{figure*}

\begin{figure*}
\centering
\begin{minipage}{175mm}
\includegraphics[width=62mm,angle=-90]{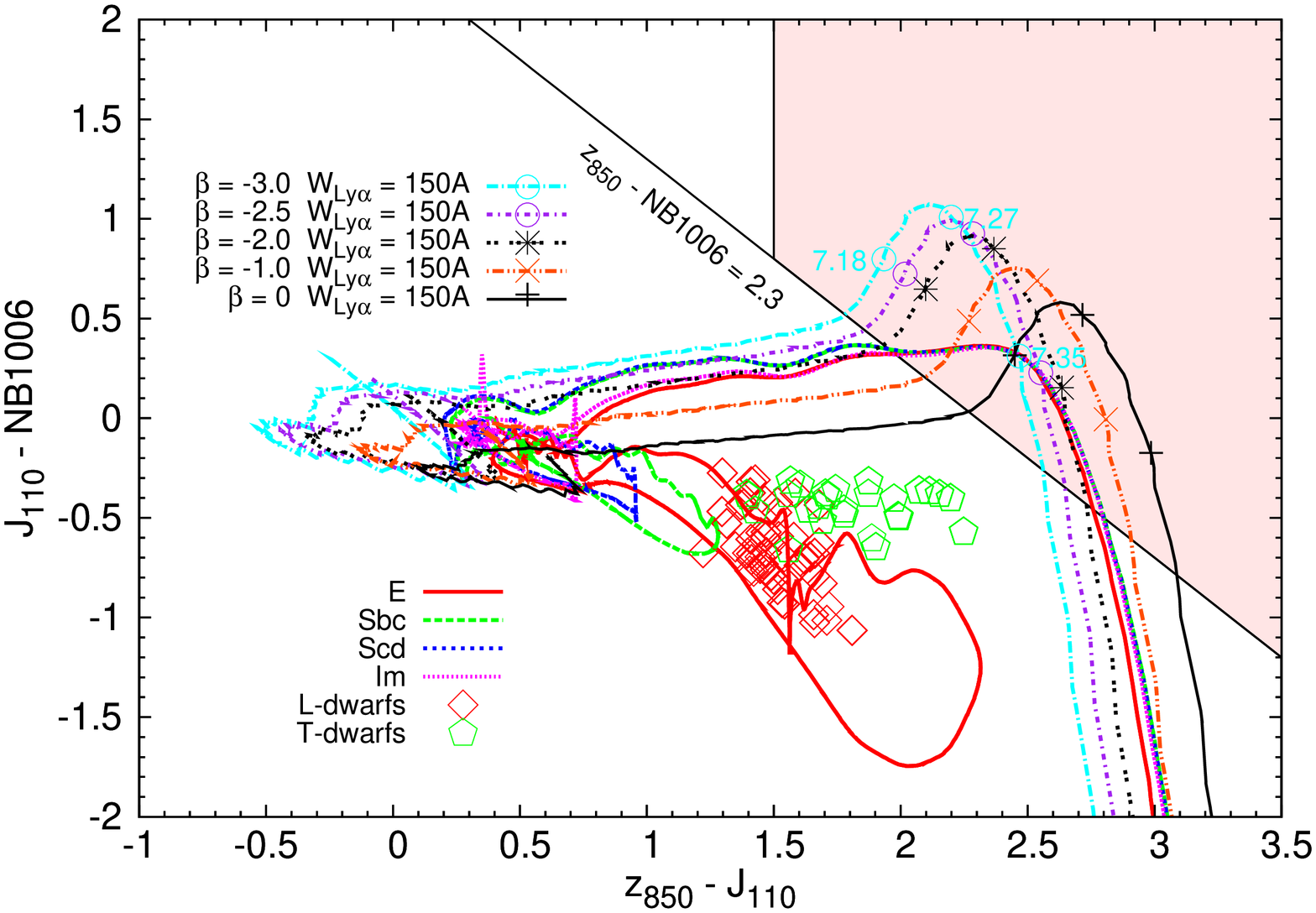}
\includegraphics[width=62mm,angle=-90]{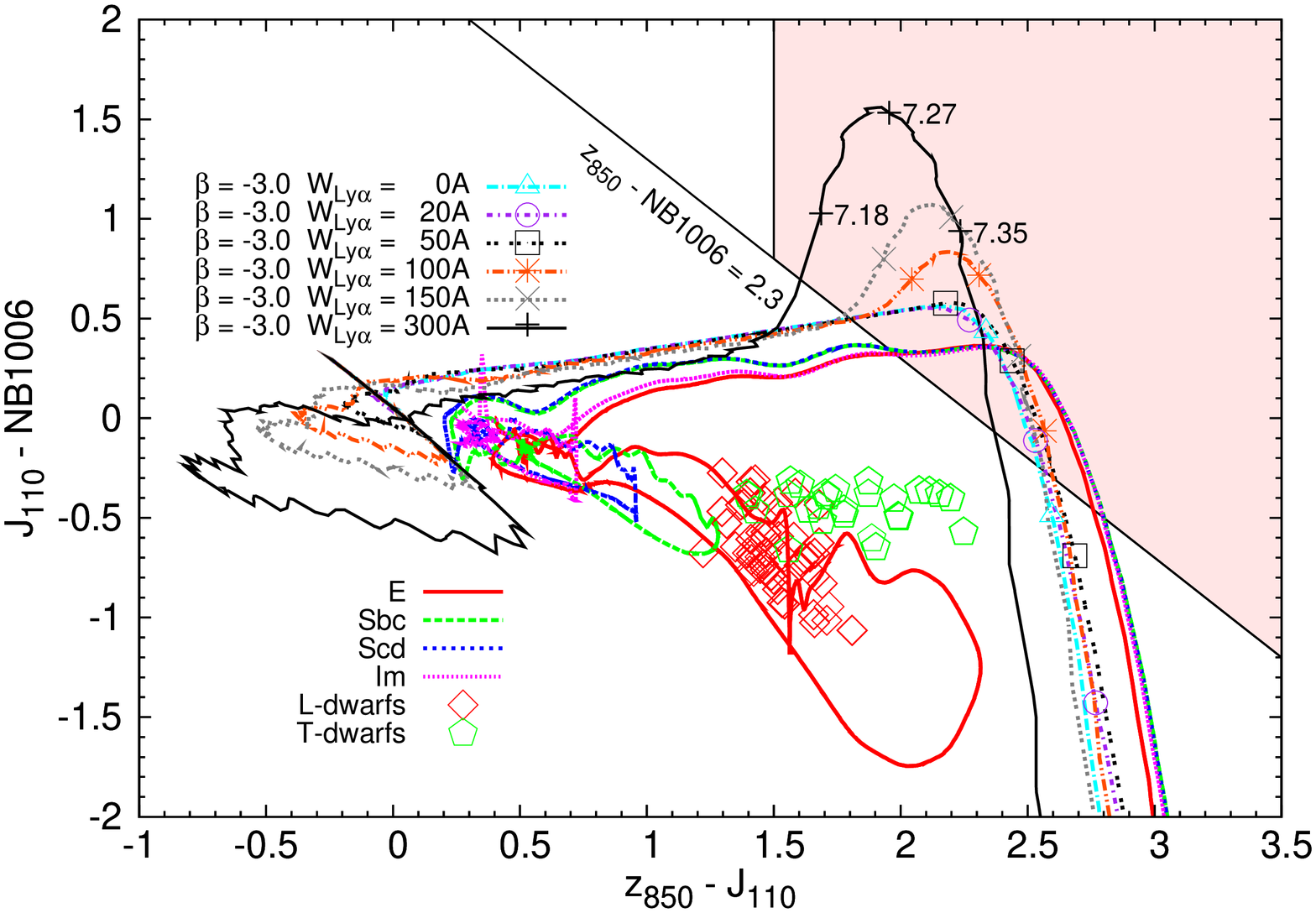}
\caption{The two color diagram, $J_{110} - {\rm NB1006}$ versus $z_{850} - J_{110}$, of various types of galaxies. The colors of E (elliptical), Sbc, Scd and Im (irregular) galaxies are calculated by using \citet{CWW80} template spectra. The color of an LAE is calculated by using the spectrum $f_{\lambda} \propto \lambda^{\beta}$ for a variety of $\beta$ and $W_{{\rm Ly}\alpha}^{\rm rest}$. The symbols indicate the locations of the minimum, central and maximum redshifts ($z=7.18$, 7.27 and 7.35 from left to right) of Ly$\alpha$ targeted by the NB1006 filter (the redshifts are labeled as a example for the $\beta=-3$, $W_{{\rm Ly}\alpha}^{\rm rest}=150$\AA~LAE and the $\beta=-3$, $W_{{\rm Ly}\alpha}^{\rm rest}=300$\AA~LAE in the left and right diagrams, respectively). The shaded region shows our LAE selection criteria (3) and (4) (See Section \ref{SelectionCriteria}). The solid diagonal line corresponds to $(z_{850} - J_{110}) - ({\rm NB1006} - J_{110}) = z_{850} - {\rm NB1006} = 2.3$, above which objects satisfy the selection criterion (3). Also, colors of L1--L9.5 and T0.5--T8 type dwarf stars are calculated by using their spectra and plotted (see the text in Section \ref{SelectionCriteria} for the details of dwarf data). ({\it Left}) Colors of LAEs with $W_{{\rm Ly}\alpha}^{\rm rest}=150$\AA~are plotted to see the dependency of the colors on $\beta$. Actually, the colors strongly depend on the $\beta$. ({\it Right}) Colors of an LBG and LAEs with $\beta=-3$ are plotted as a representative, as recent study of \citet{Ono10} suggests that the $z\sim7$ LBGs ($z_{850}$-dropouts) and $z=5.7$ and $z=6.6$ LAEs have the continuum slope of $\beta\simeq-3$. The criteria (3) and (4) we adopt in Section 3.3 select $z \sim 7.2$--7.4 LAEs with $W_{{\rm Ly}\alpha}^{\rm rest} > 50$\AA, $z\sim7.2$--7.3 LAEs with $W_{{\rm Ly}\alpha}^{\rm rest}=20$--50\AA~and $z\sim7$--7.2 LBGs with $W_{{\rm Ly}\alpha}^{\rm rest}=0$.}
\label{CandidateColor2}
\end{minipage}
\end{figure*}

\begin{figure*}
\centering
\begin{minipage}{175mm}
\includegraphics[width=62mm,angle=-90]{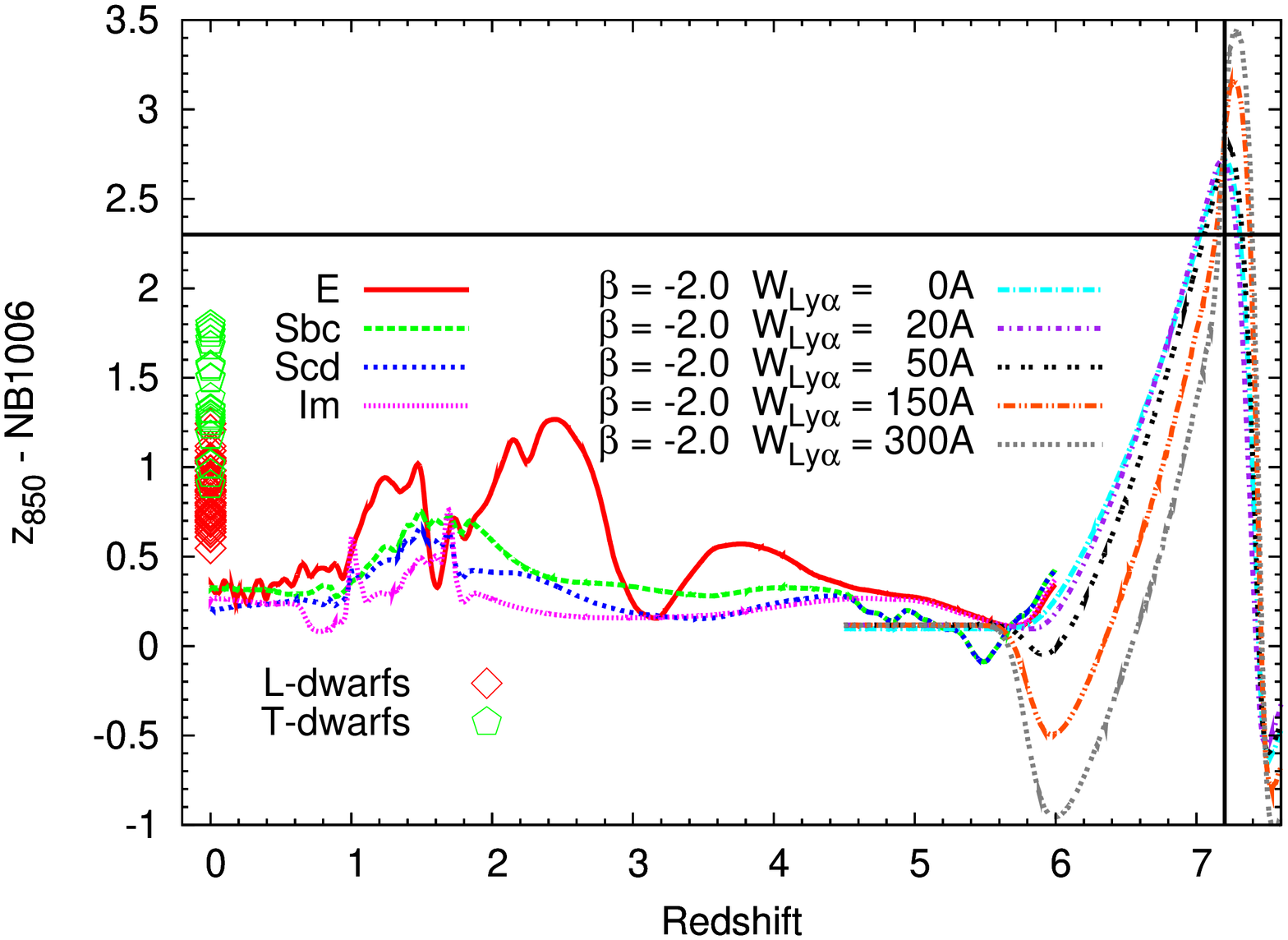}
\includegraphics[width=62mm,angle=-90]{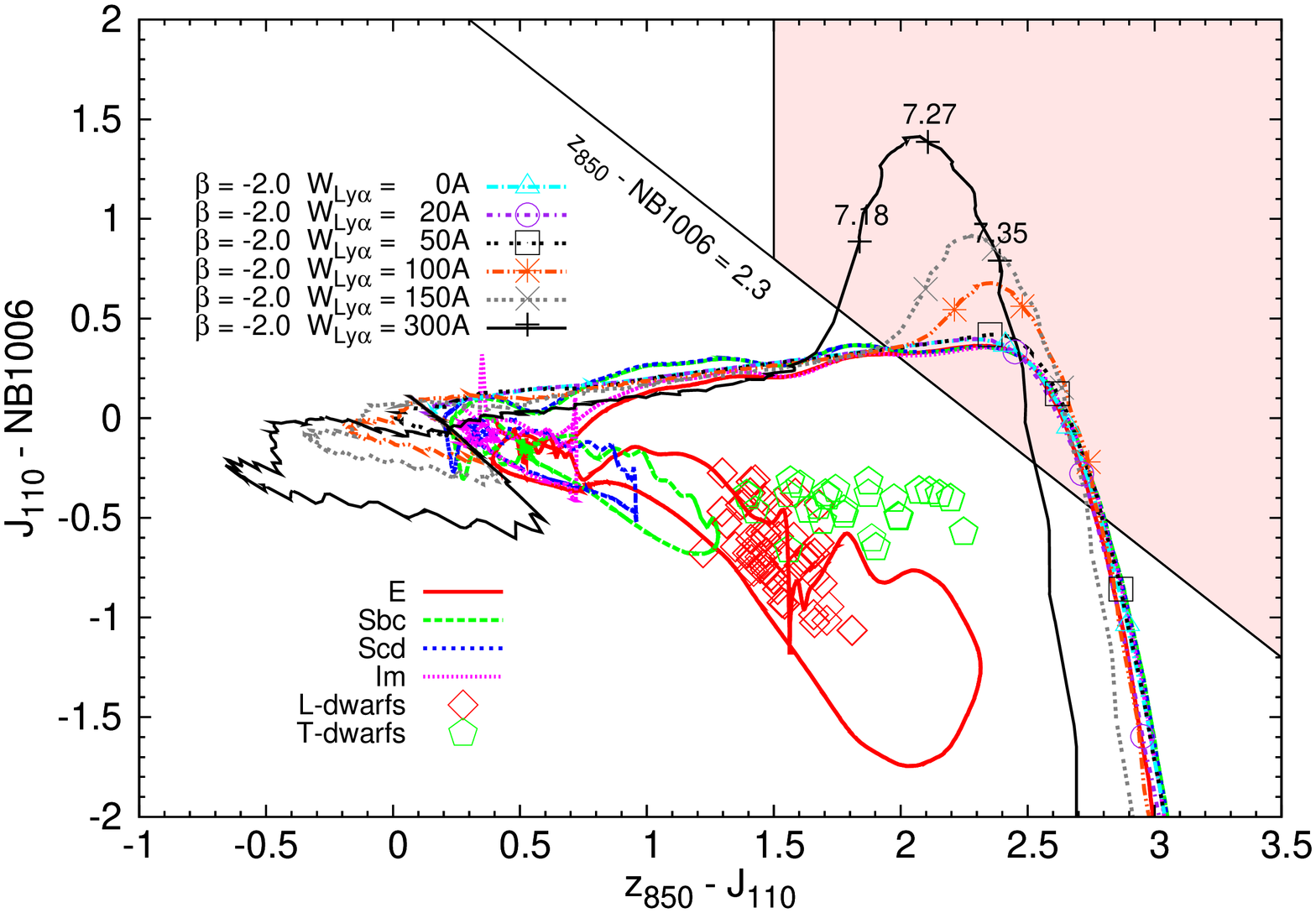}
\caption{The same figures as the right panels in Figures \ref{CandidateColor1} and \ref{CandidateColor2} but for $\beta=-2$. We plot these to see how the colors of LAEs and LBGs change from the case of $\beta=-3$ because more recent studies such as \citet{Wilkins11} and \citet{Dunlop12} show that $\beta = -2$ for $z_{850}$-dropouts. The $z_{850}-J_{110}$ color is more severe and $J_{110}-{\rm NB1006}$ is more relaxed than the case of $\beta=-3$. Similarly to the case of $\beta=-3$ (see the right panel of Figure \ref{CandidateColor2}), the criteria (3) $z_{850} - {\rm NB1006} > 2.3$ and (4) $z_{850} - J_{110} > 1.5$ we adopt in Section 3.3 also allow to select $z \sim 7.2$--7.4 LAEs with $W_{{\rm Ly}\alpha}^{\rm rest} > 50$\AA, $z\sim7.2$--7.3 LAEs with $W_{{\rm Ly}\alpha}^{\rm rest}=20$--50\AA~and $z\sim7$--7.3 LBGs with $W_{{\rm Ly}\alpha}^{\rm rest}=0$ for $\beta=-2$.}
\label{CandidateColor3}
\end{minipage}
\end{figure*}

\subsection{LAE Selection Criteria\label{SelectionCriteria}}
Figure \ref{FilterTransmission} shows that the bandpass of NB1006 is located at the red side of the bandpass of ACS $z_{850}$ filter as well as in the middle of the bandpass of NICMOS $J_{110}$ filter. If the Ly$\alpha$ emission is redshifted in the bandpass of NB1006, the LAE is expected to show significant excess in NB1006 with respect to $z_{850}$ and $J_{110}$. This characteristic is used to isolate $z=7.3$ LAEs from other objects. We investigated the expected $z_{850} - {\rm NB1006}$ and $J_{110} - {\rm NB1006}$ colors for $z=7.3$ LAEs and derived candidate selection criteria. We created model spectra of LAEs by assuming the power law continuum $f_{\lambda} \propto \lambda^{\beta}$ with several different slopes $\beta=-3, -2.5, -2, -1.5, -1, 0$ and adding them the Ly$\alpha$ emission with rest frame equivalent width of $W_{\rm Ly\alpha}^{\rm rest}=0, 20, 50, 100, 150$ and 300\AA. We did not assume any specific line profile or velocity dispersion of Ly$\alpha$ emission. Instead, we simply added the total line flux value to the spectra at 1216\AA. Then, the spectra were redshifted to $z=0$--8, and Ly$\alpha$ absorption by intergalactic medium (IGM) was applied to them, using the prescription of \citet{Madau95}. 

Colors of these model LAEs were calculated using their redshifted spectra and transmission curves of $z_{850}$, NB1006 and $J_{110}$ filters and plotted as a function of redshift in Figures \ref{CandidateColor1}, \ref{CandidateColor2} and \ref{CandidateColor3}. For comparison, we also calculated the colors of E (elliptical), Sbc, Scd and Im (irregular) galaxies by using the \citet{CWW80} template spectra and plotted in Figures \ref{CandidateColor1}, \ref{CandidateColor2} and \ref{CandidateColor3}. Moreover, we also plot the colors of L- and T-type dwarf stars, which can be interlopers, in Figures \ref{CandidateColor1}, \ref{CandidateColor2} and \ref{CandidateColor3}. We calculate the colors by using the filter transmission curves and the spectra of L1--L9.5 and T0.5--T8 dwarfs \citep{Burgasser00,Burgasser03,Chiu05,Chiu06,Cruz03,Fan00,Geballe96,Geballe01,Geballe02,Golimowski98,Kirkpatrick97,Kirkpatrick99,Kirkpatrick00,Knapp04,Leggett99,Leggett00,Leggett01,Leggett02a,Leggett02b,Martin99,McLean03,Reid00,Reid01,Ruiz97,Schultz98,Strauss99,Tsvetanov00} compiled by S. Leggett in her L and T dwarf data archive\footnote{http://staff.gemini.edu/\~{}sleggett/LTdata.html.}.

As clearly seen, an LAE is expected to produce significant flux excess in NB1006 against $z_{850}$ and $J_{110}$ at around $z=7.3$. The model spectra without Ly$\alpha$ emission ($W_{\rm Ly\alpha}^{\rm rest}=0$), which are considered spectra of LBGs, also show some extent of NB1006 excess at $z\sim7$--7.3. Meanwhile, elliptical galaxies at $z \sim 1$--3 also show some excess of $z_{850} - {\rm NB1006} \sim 0.9$--1.3 due to their 4000\AA~Balmer break. Also, L and T dwarfs display excess of $z_{850} - {\rm NB1006} \sim 0.5$--1.9 due to their spectral shapes. Hence, we adopt $z_{850} - {\rm NB1006} > 2.3$ as an NB1006-excess criterion that selects $z\sim7.2$--7.4 LAEs with $W_{\rm Ly\alpha}^{\rm rest} > 50$\AA~and $z\sim7.2$--7.3 LAEs with $W_{\rm Ly\alpha}^{\rm rest} = 20$--50\AA~and LBGs ($W_{\rm Ly\alpha}^{\rm rest} = 0$) but avoids contamination from the ellipticals and dwarfs. \citet{Richard08} and \citet{Bouwens09} conducted surveys of $z_{850}$-dropout LBGs in Abell 2390 and CL 0024 and detected two and one LBG candidates, respectively. Comparison with them can distinguish between LAEs and LBGs in our candidate sample. We use the following criteria to select LAEs: 
\begin{equation}
\bf   {\rm \bf NB1006} \ge 4\sigma
\end{equation}
\begin{equation}
   {\rm Optical~bands} < 2\sigma~{\rm and~no~object~is~seen~visually}
\label{OpticalNullDetections}
\end{equation}
\begin{equation}
\bf   z_{850} - {\rm \bf NB1006} > 2.3
\label{NBexcess1}
\end{equation} 
\begin{equation}
\bf   z_{850} - J_{110} > 1.5~{\rm \bf (where}~J_{110}~{\rm \bf images~are~available)}
\label{NBexcess2}
\end{equation} 
We set the criterion (1) as our object detection limit. This keeps the objects brighter than the $4\sigma$ limiting magnitudes ($2\farcs0$ aperture), ${\rm NB1006}=23.47$ and 23.39 for Abell 2390 and CL 0024, respectively. The criterion (2) means non-detections ($<2\sigma$) in $V_{555}$ and $I_{814}$ for Abell 2390 and $g_{475}$, $r_{625}$ and $i_{775}$ for CL 0024 (these are all measured in total magnitudes). We also visually inspect the optical bands to ensure that no object is seen in each image. This criterion is applied since the flux of an LAE shortward of Ly$\alpha$ emission should be absorbed by IGM and can help reduce the number of interlopers such as lower redshift galaxies with other types of emission lines (e.g., H$\beta$, [OIII], [OII] or H$\alpha$ emitters). However, we cannot completely remove all the interlopers with this criterion, as spectroscopy of $z=5.7$ and 6.6 LAEs conducted by \citet{Shimasaku06} and \citet{Kashikawa11} who also applied similar criteria to reduce foreground interlopers, for example, identified a few of their LAE candidates as low redshift line emitters (at least 6\% and 2\% in their $z=5.7$ and $z=6.6$ samples, respectively).   

The criterion (3) is the NB1006 excess against $z_{850}$ due to the $z\sim7.3$ Ly$\alpha$ emission. The criterion (4) $z_{850} - J_{110} > 1.5$ is to be applied to the objects at the positions in the NB1006 image where the $J_{110}$-band image is available to reduce the contamination from low-$z$ line emitters. The colors $z_{850}-{\rm NB1006}$ and $z_{850}-J_{110}$ of the criteria (3) and (4) are measured with a $2\farcs0$ and $0\farcs6$ apertures. In the calculation of the colors, if the $z_{850}$ and $J_{110}$ magnitudes are fainter than $1\sigma$, they are replaced by the $1\sigma$ values. In the right panels of Figures \ref{CandidateColor1} and \ref{CandidateColor2} and the both panels in Figure \ref{CandidateColor3}, we plot colors of LAEs/LBGs with $\beta=-3$ and $\beta=-2$ to see the difference as recent study of \citet{Ono10} suggests that the $z\sim7$ LBGs ($z_{850}$-dropouts) and $z=5.7$--6.6 LAEs have the continuum slope of $\beta\simeq-3$ while more recent studies such as \citet{Wilkins11} and \citet{Dunlop12} show that $\beta=-2$ for $z_{850}$-dropouts. The $z_{850}-J_{110}$ color is more severe and $J_{110}-{\rm NB1006}$ is more relaxed in the case of $\beta=-2$ than $\beta=-3$. The criteria (3) and (4) allow to select $z \sim 7.2$--7.4 LAEs with $W_{{\rm Ly}\alpha}^{\rm rest} > 50$\AA, $z\sim7.2$--7.3 LAEs with $W_{{\rm Ly}\alpha}^{\rm rest}=20$--50\AA~and $z\sim7$--7.2 or 7.3 LBGs with $W_{{\rm Ly}\alpha}^{\rm rest}=0$ for both $\beta=-3$ and $\beta=-2$.

\subsection{Candidate Selection\label{CandidateSelection}}
By applying the LAE selection criteria to the NB1006-detected object catalogs, we found 5 and 17 objects satisfying the criteria in Abell 2390 and CL 0024, respectively. Then, we visually inspected their NB1006 images and removed obviously spurious objects such as tails of saturated pixels from bright stars, halos of bright stars and cosmic rays with extremely long narrow shapes. As a result, we were left with two objects in Abell 2390 and three in CL 0024. The two objects in Abell 2390 and two out of the three in CL 0024 are detected only in NB1006 and not detected in any other bands at wavelength longer than NB1006. Two out of them are located outside of the $J_{110}$ and $H_{160}$ images of Abell 2390 and CL 0024, and we do not know whether they would be detected if they were observed in these bands. Hence, there is a possibility that these objects might be noises. Meanwhile, one object in CL 0024 is detected in the $K$, [3.6] and [4.8]. However, this does not necessarily guarantee that the detection in NB1006 is real. This object is also located outside of the $J_{110}$ and $H_{160}$ images of CL 0024, and we do not know whether it would be detected if it were observed in these bands. 

Therefore, to check if the five detections in NB1006 are real, we visually inspected all the dithered 20 minutes exposure frames of these objects before they were stacked to obtain the final images. We found that at least two out of all the 20 minutes frames had cosmic rays with narrow elongated shapes and high fluxes at the positions of all the five objects. These cosmic rays overlapped and were not completely removed at the time of the stacking, which caused the detections in NB1006. Hence, we conclude that all the five objects that satisfied the LAE criteria are not LAEs but the residuals of cosmic rays. Usually, cosmic rays can be cleanly removed when stacking the dithered exposure frames. However, in our case this time, the number of dithered exposure frames is not large enough. Hence, if two or more cosmic rays accidentally hit the same position in different exposure frames and overlap with each other at the time of stacking, they cannot be removed completely. Eventually, we could not detect any $z=7.3$ LAE candidates behind Abell 2390 and CL 0024.

\begin{figure}
\includegraphics[width=62mm,angle=-90]{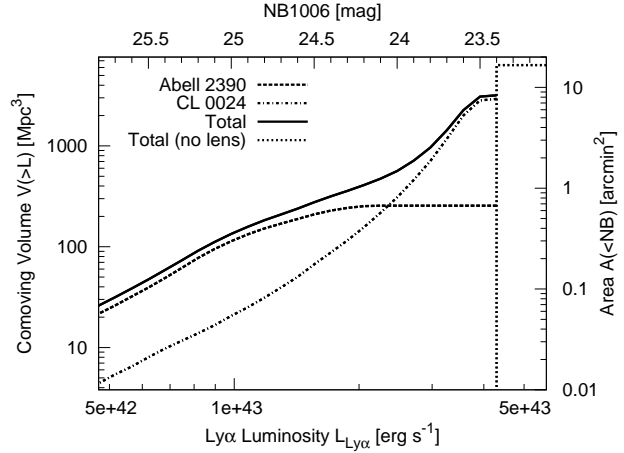}
\caption{Unlensed comoving volume (surface area) sampled with the NB1006 filter brighter than a given unlensed Ly$\alpha$ luminosity (magnitude) in the source plane for each cluster and for total of the two clusters. The dotted line shows the total volume (area) sampled in the absence of lensing. The limiting Ly$\alpha$ luminosity shown here is calculated assuming the conversion factor of 0.7 between the NB1006 total flux and the Ly$\alpha$ flux, $F_{{\rm Ly}\alpha} = 0.7 \times F_{\rm NB}$, in the case of $z=7.3$ and $W_{{\rm Ly}\alpha}^{\rm rest}=20$\AA~(the Ly$\alpha$ line threshold of our LAE candidate selection criteria).}
\label{VolumeLLya}
\end{figure}

\subsection{Survey Depth and Volume \label{DepthVolume}}
To understand how the depth improved and volume decreased in our survey with the presence of lensing magnification (the trade-off between depth and volume) and to what depth and within what area we could not detect $z=7.3$ LAEs, we calculated the unlensed comoving volume sampled with the NB1006 filter as a function of limiting unlensed Ly$\alpha$ luminosity $L_{{\rm Ly}\alpha}$ in the source plane for each cluster. This can be derived by calculating the unlensed surface area as a function of magnification $\mu$ (in the unit of mag) and converting the area and the NB1006 magnitude unlensed with $\mu$ to the comoving volume and the Ly$\alpha$ luminosity, respectively. The area as a function of $\mu$ is calculated assuming $z=7.3$ and using the same lensing model used in \citet{Richard08}. This calculation is made in the field of views (FOVs) of the WFPC2 and ACS for Abell 2390 and CL 0024, respectively, because the surveyed areas in the image planes of Abell 2390 and CL 0024 are limited by the WFPC2 and ACS images used for the LAE selection criterion (\ref{OpticalNullDetections}) (null detections in the bands blueward of Ly$\alpha$) combined with the criterion (3). The conversion of the area to the volume is done by considering the comoving distance along the line of sight corresponding to the bandwidth (FWHM) of the NB1006 filter. 

\begin{figure}
\includegraphics[width=62mm,angle=-90]{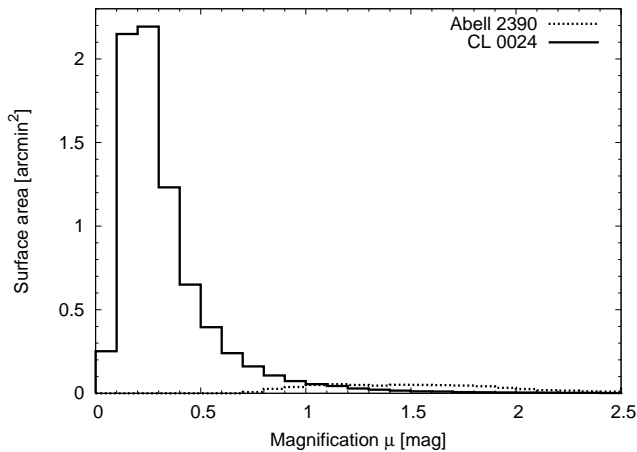}
\caption{Distribution of magnification factors $\mu$ (in magnitudes) calculated by using the lensing models of \citet{Richard08} and assuming a point source at $z=7.3$. The horizontal axis is $\mu$ in 0.1 mag bin while the vertical axis shows unlensed surface area.}
\label{MagDistribution}
\end{figure}

When calculating the Ly$\alpha$ luminosity from the NB1006 magnitude, we used the conversion factor of 0.7. That is, we consider that 70\% of the flux corresponding to the NB1006 magnitude comes from Ly$\alpha$ emission. This conversion factor is derived as follows. We assume that the NB1006 flux $F_{\rm NB}$ consists of contributions from Ly$\alpha$ and UV continuum fluxes, $F_{{\rm Ly}\alpha}$ and $F_{\rm cont}$. 
\begin{equation}
F_{\rm NB}=F_{{\rm Ly}\alpha}+F_{\rm cont} 
\label{fluxNB}
\end{equation}
The Ly$\alpha$ flux is related to the UV continuum flux density $f_{\lambda,{\rm cont}}$ with observed Ly$\alpha$ equivalent width $W_{{\rm Ly}\alpha}^{\rm obs}$.
\begin{equation}
F_{{\rm Ly}\alpha}=W_{{\rm Ly}\alpha}^{\rm obs}f_{\lambda,{\rm cont}} 
\label{fluxLya}
\end{equation}
On the other hand, the $f_{\lambda,{\rm cont}}$ is the UV continuum flux divided by the wavelength from Ly$\alpha$ emission to the edge of the NB1006 filter bandpass ($\lambda_{\rm NB}^{\rm max}-\lambda_{{\rm Ly}\alpha}^{\rm obs}$),  
\begin{equation}
f_{\lambda,{\rm cont}}=F_{\rm cont}/(\lambda_{\rm NB}^{\rm max}-\lambda_{{\rm Ly}\alpha}^{\rm obs})
\label{fluxcont}
\end{equation}
where $\lambda_{\rm NB}^{\rm max}=10160$\AA, and we assume that Ly$\alpha$ emission is at $z=7.3$, i.e., $\lambda_{{\rm Ly}\alpha}^{\rm obs} = (1+z)1216$\AA~$= 10092.8$\AA. Solving these equations (\ref{fluxNB})--(\ref{fluxcont}) for $F_{{\rm Ly}\alpha}$, we have
\begin{equation}
F_{{\rm Ly}\alpha}=F_{\rm NB}/\left(1+\frac{\lambda_{\rm NB}^{\rm max}-\lambda_{{\rm Ly}\alpha}^{\rm obs}}{W_{{\rm Ly}\alpha}^{\rm obs}}\right) \label{fluxfraction}
\end{equation} 
Because our LAE selection criterion (3) corresponds to the rest frame Ly$\alpha$ equivalent width threshold of $W_{{\rm Ly}\alpha}^{\rm rest} \ge 20$\AA, we have $W_{{\rm Ly}\alpha}^{\rm obs} = W_{{\rm Ly}\alpha}^{\rm rest}(1+z) \ge 166$\AA~and accordingly $F_{{\rm Ly}\alpha} \ge 0.7 F_{\rm NB}$. The $F_{{\rm Ly}\alpha} = 0.7 F_{\rm NB}$ or equivalently $L_{{\rm Ly}\alpha} = 0.7 L_{\rm NB}$ in luminosity corresponds to the limiting luminosity in our lensing survey and the volume should be calculated down to this limit. Figure \ref{VolumeLLya} shows the volume sampled with the NB1006 as a function of limiting Ly$\alpha$ luminosity in the source plane for each cluster and the sum of the two clusters. The trade-off of increase in depth and reduction in volume is clearly seen. In Figure \ref{MagDistribution}, we show the distribution of magnification factors $\mu$ (unlensed area versus $\mu$ in 0.1 mag bin) in each cluster. The $\mu$ ranges from 0 to 2.5 mag and the total unlensed area (sum of Abell 2390 and CL 0024) in the source plane is 8.4 arcmin$^2$, a factor of 2 reduction from the lensed area in the image plane (i.e., sum of the FOVs of the WFPC2 and ACS; see Figure \ref{NB1006images}). The unlensed $4\sigma$ limiting magnitudes in the $\mu=1.3$ (median magnification) and $\mu=2.5$ areas are NB1006 = 24.7 and 25.9. These correspond to the Ly$\alpha$ line flux limits of $F_{{\rm Ly}\alpha} \simeq 2.1 \times 10^{-17}$ and $6.9 \times 10^{-18}$ erg s$^{-1}$ cm$^{-2}$ or $L_{{\rm Ly}\alpha} \simeq 1.3 \times 10^{43}$ and $4.4 \times 10^{42}$ erg s$^{-1}$, respectively. Therefore, to the limits varying from $L_{{\rm Ly}\alpha} \simeq 4.4 \times 10^{43}$ ($\mu=0$) to $4.4 \times 10^{42}$erg s$^{-1}$ (as shown in Figure \ref{VolumeLLya}) within the total area of 8.4 arcmin$^2$, we could not detect any $z=7.3$ LAEs. 

\begin{table*}
\centering
\begin{minipage}{177mm}
\begin{center}
\caption{Comparison of the expected detection number of $z\sim7$ LAEs in lensing and blank field surveys using an 8m class ground-based telescope, red-sensitive CCDs and a narrowband filter.}
\label{LAEnumbers}
\begin{tabular}{@{}cccccccccccccc@{}}
\hline 
 & \multicolumn{4}{c}{Case (1)} & \multicolumn{3}{c}{Case (2)} & \multicolumn{3}{c}{Case (3)} & \multicolumn{3}{c}{Case (4)}\\ 
Obs.  & \multicolumn{4}{c}{\underline{Observing many lensing clusters}} & \multicolumn{3}{c}{\underline{Observing 1 lensing cluster}} & \multicolumn{3}{c}{\underline{Observing 1 blank field}} & \multicolumn{3}{c}{\underline{Observing 1 blank field}}\\ 
Time$^a$ & Number of & \multicolumn{3}{c}{\underline{\#LAEs in ACS FOV$^c$}} & \multicolumn{3}{c}{\underline{\#LAEs in ACS FOV$^c$}} & \multicolumn{3}{c}{\underline{\#LAEs in ACS FOV$^c$}} & \multicolumn{3}{c}{\underline{\#LAEs in S-Cam FOV$^c$}}\\   
(hour) & Clusters$^b$ & O10 & H11 & H12 & O10 & H11 &H12 & O10 & H11 & H12 & O10 & H11 & H12\\ 
\hline 
2   &   1 & 0.02 & 0.22 & 0.03 & 0.02 & 0.22 & 0.03 & 0.01 & 0.02 & 0.01 & 0.86 & 1.5  & 0.47\\ 
4   &   2 & 0.04 & 0.45 & 0.06 & 0.04 & 0.61 & 0.07 & 0.02 & 0.14 & 0.03 & 1.73 & 9.9  & 2.0\\ 
10  &   5 & 0.11 & 1.1  & 0.15 & 0.11 & 2.1 & 0.21 & 0.06 & 0.87 & 0.12 & 4.38 & 63.7 & 8.7\\ 
20  &  10 & 0.21 & 2.2  & 0.30 & 0.21 & 4.4 & 0.40 & 0.12 & 2.5  & 0.28 & 8.83 & 183  & 20.3\\ 
100 &  50 & 1.1  & 11.2 & 1.5  & 1.1 & 17.4 & 1.3 & 0.62 & 14.4 & 1.2  & 45.0 & 1048 & 85.9\\ 
200 & 100 & 2.1  & 22.4 & 3.0  & 2.1 & 27.3 & 1.9 & 1.2  & 24.9 & 1.9  & 90.7 & 1819 & 138\\
\hline
\multicolumn{2}{c}{\#LAEs per hour$^d$} & 0.01 & 0.11 & 0.02 & 0.01$^f$ & 0.22$^f$ & 0.02$^f$ & 0.01$^f$ & 0.13$^f$ & 0.01$^f$ & 0.44$^f$ & 9.17$^f$ & 1.02$^f$\\
\multicolumn{2}{c}{Hours per LAE$^e$}   & 95   & 8.9  & 66   & 96$^f$ & 4.6$^f$ & 50$^f$ & 165$^f$  & 8.0$^f$  & 72$^f$   & 2$^f$    & 0.1$^f$ & 1.0$^f$\\
\hline
\end{tabular}
\end{center}
NOTE: This is a model case that an 8m class telescope = Subaru, red-sensitive CCDs = Suprime-Cam fully depleted CCDs and a narrowband filter = NB973 ($\lambda_c=9755$\AA, ${\rm FWHM}=200$\AA).
$^a$Total observing time in the unit of hour. In the Case (1), we considered the integration time per cluster of 2 hours, and hence total observing time $=$ 2 hours $\times$ the number of clusters. In the Cases (2)--(4), the total observing time means integration time of one cluster or one blank field.
$^b$The number of lensing clusters observed for the Case (1). 
$^c$The number of LAEs expected to be detected in the field of view (FOV) of the ACS and the Suprime-Cam estimated by extrapolating the $z=7$ Ly$\alpha$ LF derived by O10 \citep{Ota10} and integrating the best-fit \citet{Schechter76} Ly$\alpha$ LFs derived by H11 \citep{Hibon11} and H12 \citep{Hibon12}. In the case of H12, we use their "common sample LF" listed in their Table 4. Because $\log L^*= 43.71$ listed in their Table 4 was found to be a typo, we use the correct one, $\log L^*= 42.71$ (P. Hibon 2012, private communication) for our calculation.   
$^d$The number of LAEs expected to be detected in one hour. 
$^e$The time (in hour) necessary to detect one LAE. 
$^f$The values in the case of integration time of 20 hours (a practical observing time for an 8m class ground-based telescope) are listed because the expected detection number of LAEs varies as slope of the Ly$\alpha$ LF changes from the bright end to the survey depths, and the survey depth depends on the integration time.    
\end{minipage}
\end{table*}

\section{Discussion}
\subsection{Comparison with Blank Field Surveys\label{LF}}
To understand the meaning of the null detection of $z=7.3$ LAEs in our lensing survey, we estimate the expected detection numbers of $z\sim7$ LAEs in lensing surveys and compare them with those in blank field surveys in the case of spending the same amount of observing time with an 8m class ground based telescope, a camera with CCDs sensitive to $z\sim7$ Ly$\alpha$ (red-sensitive CCDs), and a narrowband filter. With this estimation and comparison, we can see if the null detection in our lensing survey was an anticipated result and know what is the most efficient way to detect $z\sim7$ LAEs. We consider a model case that an 8m class telescope = Subaru Telescope, red-sensitive CCDs = Suprime-Cam's fully depleted CCDs, a narrowband filter = NB973 ($\lambda_c=9755$\AA, FWHM $=$ 200\AA) and a lensing cluster = CL 0024. The NB973 is the filter used by \citet{Ota08,Ota10} and \citet{Hibon12} for their $z=7$ LAE surveys.

We first calculate the expected detection number of LAEs in lensing and field surveys by using Ly$\alpha$ LFs derived from the several different narrowband $z=7$ LAE blank field surveys conducted by \citet{Ota10}, \citet{Hibon11} and \citet{Hibon12} (hereafter O10, H11 and H12) and then consider the case for the narrowband NB1006 $z=7.3$ LAE lensing survey (the current study). We use the three different $z=7$ LFs for the calculation since we do not have any $z=7.3$ LFs and we also want to see how the expected number of LAEs changes with different LFs. 

We compare four cases (see also Table \ref{LAEnumbers}): Case (1) a survey whose one pointing covers the area comparable to the {\it HST} ACS field of view (FOV $\sim 11.3$ arcmin$^2$ and corresponds to the area just covering the size of a lensing cluster CL 0024) and observing many clusters (1, 2, 5, 10, 50 and $100\times$ CL 0024), Case (2) a lensing survey covering an area with the {\it HST} ACS FOV and observing only one cluster (CL 0024) to several different depths, Case (3) a survey observing one blank field covering an area with the ACS FOV to several different depths and Case (4) a survey observing one blank field covering an area comparable to the Subaru Suprime-Cam FOV to several different depths. For the Case (1), we consider integration time per cluster of 2 hours as a model case, because spending short integration time for each cluster by counting on the magnification and observing many clusters are considered a practical observation strategy with a ground based telescope. For the Cases (2)--(4), we consider integration times of 2, 4, 10, 20, 100 and 200 hours, which corresponds to 2 hours $\times$ 1, 2, 5, 10, 50 and 100 clusters, to compare with the result of the Case (1).  

\subsubsection{The Expected Detection Number of $z\sim7$ LAEs in Lensing Surveys\label{LAEinLensing}}
 We make the calculation of this with the following 5 steps. (1) First, we convert the magnification $\mu$ distribution of CL 0024 in Figure \ref{MagDistribution} to unlensed comoving volume $V(L)$ sampled down to each unlensed limiting Ly$\alpha$ luminosity ($L_{{\rm Ly}\alpha}$) bin.  In this procedure, we convert each $\mu$ to unlensed $5\sigma$ NB973 limiting magnitude with $m^{\rm lens}_{\rm lim}$ + $\mu$ and then to the corresponding $L_{{\rm Ly}\alpha}$, where the $m^{\rm lens}_{\rm lim}$ is lensed $5\sigma$ NB973 limiting magnitudes reached by the integration times we consider. We use the conversion factor used by O10 to convert NB973 magnitudes to $L_{{\rm Ly}\alpha}$ for the O10 case and the conversion factor used by H11 for the H11 and H12 cases. (2) Next, we calculate cumulative LAE number density $n($$>$$L)$ down to each $L_{{\rm Ly}\alpha}$ bin by linearly extrapolating the $z=7$ Ly$\alpha$ LF ($\log n($$>$$L)$ vs. $\log L_{{\rm Ly}\alpha}$) obtained by O10. This LF has a slope of $-d(\log n)/d(\log L_{{\rm Ly}\alpha}) \simeq 1.5$. Meanwhile, to obtain $n($$>$$L)$'s for H11 and H12, we integrate the best-fit \citet{Schechter76} $z=7$ Ly$\alpha$ LFs derived by them to each $L_{{\rm Ly}\alpha}$ bin (See the footnote of Table \ref{LAEnumbers} for more details). (3) By multiplying each $V(L)$ obtained in (1) by $n($$>$$L)$ obtained in (2), we calculate the number of LAEs $N(L)=V(L)$$n($$>$$L)$ expected to be detected down to each $L_{{\rm Ly}\alpha}$ bin. (4) Summing up all the $N(L)$ in all the $L_{{\rm Ly}\alpha}$ bin, we obtain the number of LAEs expected to be detected behind one lensing cluster (in this case, CL 0024). (5) The number of LAEs expected to be detected in surveying more than one clusters is calculated by simply multiplying the number of LAEs obtained in (4) by the number of clusters in consideration.

\subsubsection{The Expected Detection Number of $z\sim7$ LAEs in Blank Field Surveys\label{LAEinField}}
We make the calculation of this with the following 3 steps. (1) Linearly extrapolating the O10's $z=7$ Ly$\alpha$ LF or integrating the H11's and H12's best-fit Schechter $z=7$ Ly$\alpha$ LFs, we calculate the cumulative LAE number densities $n($$>$$L)$'s down to the limiting $L_{{\rm Ly}\alpha}$'s corresponding to $5\sigma$ limiting NB973 magnitudes reached by the integration times we consider. (2) Multiplying these $n($$>$$L)$'s by the volume sampled by the Suprime-Cam + NB973 filter, $3 \times 10^5$ Mpc$^3$ (corresponding to an area of 824 arcmin$^2$) taken from O10, we obtain the number of LAEs expected to be detected in the blank field survey with the Suprime-Cam FOV. (3) Then, we calculate the expected detection number of LAEs in the blank field survey with the same sampled area as the lensing surveys. In the case of CL 0024, the area corresponds to the FOV of ACS $\sim11.3$ arcmin$^2$. Hence, we obtain the number of LAEs by multiplying  the number of LAEs in the Suprime-Cam blank field survey by a factor of $11.3/824$.

\subsubsection{Comparison of Lensing and Field LAE Surveys}
The result of above calculations in Sections \ref{LAEinLensing} and \ref{LAEinField} are presented in Table \ref{LAEnumbers}. First, we compare the Cases (1)--(3) for each LF. In the case of the O10 LF, the survey efficiencies of the Cases (1) and (2) are the same and better than that of the Case (3). This is expected because the O10 LF has the bright end slope of $-d(\log n)/d(\log L_{{\rm Ly}\alpha}) \simeq 1.5 > 1$ and we simply linearly extrapolate this LF to the fainter limits corresponding to the integration times to calculate the number of LAEs. In the case of the H11 LF, the Case (2) is more efficient than the Cases (1) and (3), and the Case (1) is more efficient than the Case (3) until the Case (1) observes 10 clusters and the Case (3) integrates for 20 hours; thereafter, the trend is reversed for the longer observing times. Meanwhile, in the case of the H12 LF, the Case (2) is more efficient than the Cases (1) and (3) until the observing time reaches 20 hours; thereafter, the efficiencies are comparable between the Cases (2) and (3), and that of the Case (1) is better than them.

In summary, for the observing times shorter than 20 hours, the Case (2) where we concentrate on one lensing cluster is more efficient than observing many clusters with short integration (Case (1)) or observing one blank field with an area comparable to size of the lensing cluster (Case (3)), but the trend for the longer observing times varies with LFs. All the three LFs considered here have their observed data points only at the bright end (i.e., a few $z=7$ LAE candidates with $L_{{\rm Ly}\alpha}\sim0.9$--$1\times 10^{43}$ erg s$^{-1}$), and we extrapolate such LFs linearly or with the Schechter function. This results in larger uncertainties in the slopes of the LFs at the fainter luminosity limits reached by observing times we consider. Hence, the results in Table \ref{LAEnumbers} for the shorter observing times reflect the observed data points of each LF better than those for longer observing times. In reality, spending more than 20 hours for an observation with a ground based 8m telescope is very expensive and not practical. Hence, concentrating on one lensing cluster seem to be most efficient for detecting $z=7$ LAEs, but two (O10 and H12) out of the three LFs imply that even spending 20 hours, we detect $\sim0.2$--0.4 LAEs (null detection) while the other one LF (H11) predicts 4.4 LAEs.

Next, we compare this result with the Case (4) where we observe one blank field whose area is much larger than the size of a lensing cluster (i.e., the Suprime-Cam FOV in the current case). In the Case (4), the three LFs predict the detections of 8.8--183 LAEs for 20 hours of observing time, much larger than 0.2--4.4 LAEs in the Case (2) considered above. Also, for any observing times and all the three LFs, the Case (4) is much more efficient than the Cases (1)--(3) in detecting $z=7$ LAEs. Hence, in conclusion, observing sufficiently large sky area compared to the size of lensing clusters is the most efficient way to detect $z\sim7$ LAEs with an 8m class ground based telescope, a red-sensitive CCDs and a narrowband filter of 200\AA~width.

However, three things should be noted. First, we used the magnification ($\mu$) distribution of CL 0024 as a model case for the calculation of the expected number of LAEs. The $\mu$ distribution varies from cluster to cluster, and thus the result presented in Table \ref{LAEnumbers} and the text above might not necessarily represent the most common or the most probable case for lensing surveys of $z\sim7$ LAEs. Second, as already mentioned, the uncertainty is large for the results with longer observing times due to the large uncertainty in the faint end slopes of the LFs. Finally, all the LFs considered here consist of photometric LAE candidates. If they include some contaminations, the real slopes of the LFs are different and thus so does the result in Table \ref{LAEnumbers}. The difference in the expected number of LAEs predicted by the three LFs could come from the difference in the degree of contaminations.

\subsubsection{Expected Detection Number of $z=7.3$ LAEs in the Present Lensing Survey}
We also make the similar calculation for the current $z=7.3$ LAE lensing survey (2.0 hours for A2390 + 1.3 hours for CL 0024; lensed depth of NB1006 $\sim$ 23.4 at $4\sigma$) with the narrowband NB1006 to obtain the expected detection number of LAEs by following the same procedures in Section \ref{LAEinLensing} and assuming that the $z=7.3$ Ly$\alpha$ LF is similar to $z=7$ Ly$\alpha$ LF. We find that the expected detection number of LAEs is 0.0062, 0.081 and 0.0096 if we use the O10, H11 and H12 Ly$\alpha$ LFs. This is consistent with the null detection of LAEs in our survey. Since the Suprime-Cam sensitivity at the wavelength of $z=7.3$ Ly$\alpha$ is lower than that at $z=7$ Ly$\alpha$, detecting the $z=7.3$ LAEs in a lensing survey is more difficult.

\subsection{Comparison with $z$-dropouts Previously Detected in Abell 2390}
\begin{figure}
\includegraphics[width=85mm,angle=0]{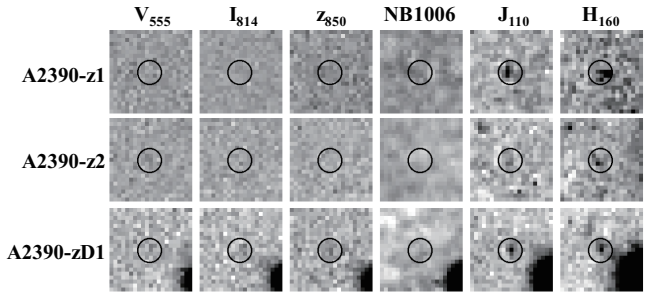}
\caption{The $4\arcsec \times 4\arcsec$ broadband and NB1006 images of $z_{850}$-dropouts previously detected in Abell 2390 by \citet{Richard08} (A2390-z1 and A2390-z2) and \citet{Bouwens09} (A2390-zD1). The A2390-z2 has a photometric redshift whose probability peaks at $z\simeq7.3$. All the three $z_{850}$-dropouts are not detected in NB1006.}
\label{VIzNBJH}
\end{figure}

Two searches for $z \ga 7$ $z_{850}$-dropout galaxies have been conducted in the lensing clusters Abell 2390 and CL 0024. \citet{Richard08} detected two $z_{850}$-dropouts, A2390-z1 and A2390-z2, while \citet{Bouwens09} detected one, A2390-zD1. Interestingly, one out of the three, the A2390-z2, has a photometric redshift whose probability peaks at $z\sim 7.3$ \citep[see Figure 7 of][]{Richard08}. Also, the magnification by the lensing cluster at the position of this object is $\mu = 1.8$ mag \citep[see Table 3 of][]{Richard08}. Hence, if this object is really at $z\sim 7.3$ and has Ly$\alpha$ emission, its line flux could be boosted by the magnification and detected in NB1006. However, we found no object detected in NB1006 at the position of the A2390-z2, as seen in Figure \ref{VIzNBJH}. There are three possible interpretations. (1) The A2390-z2 is not at $z\sim7.3$. (2) It is at $z\sim7.3$ but does not show Ly$\alpha$ emission. (3) It is at $z\sim7.3$ and have Ly$\alpha$ emission, but the emission is fainter than the NB1006 image detection limit. We cannot tell which is most likely from the current data alone. If the A2390-z2 is at $z\sim7.3$, it might be possible that its Ly$\alpha$ emission is suppressed by neutral hydrogen, as recent studies \citep{Stark10,Stark11, Ono12,Pentericci11,Schenker12} suggest that the fraction of Ly$\alpha$ emitting LBGs would be smaller at $z\ga6$--7 due to such an effect. If we assume that the A2390-z2 is at $z\sim7.3$, we estimate the unlensed upper limit on the Ly$\alpha$ line flux to be $F_{{\rm Ly}\alpha} < 4.4 \times 10^{-18}$ erg s$^{-1}$ cm$^{-2}$, by directly converting the $1\sigma$ limiting magnitude of ${\rm NB1006} = 24.98$ plus $\mu = 1.8$ mag into the flux. As seen in Figure \ref{VIzNBJH}, the A2390-z2 is detected in $H_{160}$ with $H_{160}=25.8$ mag \citep[see Table 3 of][]{Richard08}. This magnitude can be converted to the continuum flux density of $f_{\lambda,{\rm cont}} \sim 2.0\times10^{-20}$ erg s$^{-1}$ cm$^{-2}$ \AA. Using this and equation (\ref{fluxLya}) and assuming $z=7.3$, the upper limit on $F_{{\rm Ly}\alpha}$ corresponds to $W_{{\rm Ly}\alpha}^{\rm rest} < 26$\AA.

We also checked the NB1006 images of the other two objects, A2390-z1 and A2390-zD1, and confirmed no detections for the both cases (see Figure \ref{VIzNBJH}). The A2390-z1 has a photometric redshift whose probability peaks at $z\sim 6.4$ \citep[Figure 7 of][]{Richard08}, and the non-detection in NB1006 is consistent with this. On the other hand, \citet{Bouwens09} selected the A2390-zD1 as a "weak" $z_{850}$-dropout based on the same selection criteria adopted by \citet{Bouwens08}. Comparison of the colors and photometric errors of the A2390-zD1, $z_{850}-J_{110} = 1.1 \pm 0.8$ and $J_{110}-H_{160} = 0.8 \pm 0.2$ \citep[Table 2 of][]{Bouwens09}, with the $z_{850}$-dropout selection criteria in the $z_{850}-J_{110}$ vs. $J_{110}-H_{160}$ plane \citep[Figure 2 of][]{Bouwens08} suggests that this object is at the border of the $z_{850}$-dropout criteria and can be either at $z\sim 6.5$--7.0 (inside the criteria) or $z<6.5$ (outside the criteria), depending on the combination of the photometric errors in the two colors. Since these redshifts are lower than $z=7.18$, the minimum redshift of Ly$\alpha$ the NB1006 can detect, non-detection in NB1006 is a consistent result. 

\section{Conclusion}
We imaged the two lensing clusters, Abell 2390 and CL 0024, using the narrowband NB1006 (central wavelength $\sim 1005$ nm and FWHM $\sim21$ nm) with the Subaru Telescope Suprime-Cam to search for gravitationally lensed $z=7.3$ LAEs. The fully depleted CCDs of the Suprime-Cam have the sensitivity to $z\sim 7$ Ly$\alpha$ emission at $\sim 1$ $\mu$m. Also, recent studies of $z\sim 6.6$--7.7 LAEs show that the Ly$\alpha$ LFs at $z\sim7$ have the steep bright end slope of $-d(\log \phi)/d(\log L) \sim 1.5 >1$ at $L \ga L^*$, suggesting that even shallower imaging of lensing clusters could be potentially useful for detecting $z \ga 7$ bright LAEs with the gain in depth by lensing magnification possibly making up for the loss in area in the overall number of lensed LAEs to be detected. Hence, combination of the Suprime-Cam CCDs and the lensing magnification was considered to be a powerful strategy to detect $z\ga7$ LAEs. The magnification in Abell 2390 and CL 0024 varies from $\mu=0$ to 2.5 mag. We reached the lensed depth of NB1006 $\sim 23.4$ at $4\sigma$ in both lensing cluster images. Where $\mu=1.3$ (median) and $\mu=2.5$, this corresponds to the unlensed depths of NB1006 $\sim 24.7$ and 25.9 or Ly$\alpha$ line flux limits of $F_{{\rm Ly}\alpha} \simeq 2.1 \times 10^{-17}$ and $6.9 \times 10^{-18}$ erg s$^{-1}$ cm$^{-2}$, respectively.

Performing the photometry in the NB1006 images as well as deep optical to mid-infrared images of the clusters taken with the {\it HST}, Subaru and {\it Spitzer}, we constructed the NB1006-detected object catalog and applied the selection criteria of $z=7.3$ LAEs. We could not detect any LAEs to our survey limit. Using the $z=7$ Ly$\alpha$ LFs determined by three different previous surveys (O10, H11 and H12), we estimated and compared the expected detection numbers of $z\sim7$ LAEs in lensing and blank field surveys with an 8m class ground based telescope, red-sensitive CCDs and a narrowband filter. We found that a blank field survey covering an area sufficiently larger than lensing clusters is more efficient in finding $z\sim7$ LAEs than both a lensing survey observing many clusters with shallow imaging and a lensing survey imaging one cluster to a deeper luminosity limit, expected from the bright end slopes of the three $z\sim7$ Ly$\alpha$ LFs.

We also investigated the NB1006 images of the three $z\sim7$ $z$-dropouts previously detected in Abell 2390 and found that none of them are detected in the NB1006. Two of them are consistent with the predictions from the previous studies that they would be at lower redshifts. The other one has a photometric redshift of $z\simeq 7.3$, and if we assume that it is at $z=7.3$, the unlensed Ly$\alpha$ line flux would be very faint: $F_{{\rm Ly}\alpha} < 4.4 \times 10^{-18}$ erg s$^{-1}$ cm$^{-2}$ ($1\sigma$ upper limit) or correspondingly $W_{{\rm Ly}\alpha}^{\rm rest} < 26$\AA. Its Ly$\alpha$ emission might be possibly suppressed by neutral hydrogen, as recent studies suggest the smaller fraction of Ly$\alpha$ emission among LBGs at $z\ga7$ than those at lower redshifts due to such an effect. 

\section*{Acknowledgments}
This work is based in part on data collected at the Subaru Telescope, which is operated by the National Astronomical Observatory of Japan (NAOJ), observations made with the NASA/ESA {\it Hubble Space Telescope}, which is operated by the Association of Universities for Research in Astronomy, Inc., under NASA contract NAS 5-26555, and observations made with the {\it Spitzer Space Telescope}, which is operated by the Jet Propulsion Laboratory, California Institute of Technology under NASA contract 1407. This work was supported by the Grant-in-Aid for the Global COE Program "The Next Generation of Physics, Spun from Universality and Emergence" from the Ministry of Education, Culture, Sports, Science and Technology (MEXT) of Japan. We are deeply grateful to the engineers of Asahi Spectra Co., Ltd. and Hamamatsu Photonics and any people involved for developing the NB1006 filter and the fully depleted CCDs for the Suprime-Cam. These technologies made the $z = 7.3$ LAE survey possible. We express the gratitude to Masahiko Hayashi, the director of the Subaru Observatory at the time of our observation, for accepting our suggestion to observe the two lensing clusters for this study by using the gap time arising before the start of other observation. We greatly appreciate the staff at the Subaru Telescope for their kind support during our observations. We thank the anonymous referee for the useful comments that helped us to improve this paper. T.S. is supported with the fellowship from the Japan Society for the Promotion of Science. The authors recognize and acknowledge the very significant cultural role and reverence that the summit of Mauna Kea has always had within the indigenous Hawaiian community. We are most fortunate to have the opportunity to conduct observations from this mountain.



\bsp

\label{lastpage}

\end{document}